\documentclass{revtex4}
\usepackage{graphicx}
\usepackage{dcolumn}
\usepackage{graphicx,amsmath,amsfonts,amssymb}
\usepackage{bm}
\newcommand{\Zeff}{Z_{\mbox{\tiny eff}\,}}
\newcommand{\kapeff}{\kappa_{\mbox{\tiny eff}\,}}
\newcommand{\be}{\begin{equation}}
\newcommand{\ee}{\end{equation}}

\begin{document}
\title{On the macroion virial contribution to the osmotic pressure in charge-stabilized
colloidal suspensions} 
\author{E. Trizac}
\affiliation{NSF Center for Theoretical Biological Physics, 
UCSD, La Jolla, CA  92093-0374 USA}
\affiliation{LPTMS,
Univ. Paris-Sud, 
UMR 8626, Orsay F-91405 and CNRS, Orsay F-91405}
\author{Luc Belloni}
\affiliation{Laboratoire Interdisciplinaire sur l'Organisation Nanom\'etrique et 
Supramol\'eculaire          
CEA/SACLAY, Service de Chimie Mol\'eculaire, 91191 Gif sur Yvette Cedex, France }
\author{J. Dobnikar}
\affiliation{Institut f\"ur Chemie, Karl-Franzens-Universit\"at,
Heinrichstrasse 28, 8010 Graz, Austria}
\affiliation{Jozef Stefan Institute, Jamova 39, 1000 Ljubljana, Slovenia}
\author{H.H. von Gr\"unberg}
\affiliation{Institut f\"ur Chemie, Karl-Franzens-Universit\"at,
Heinrichstrasse 28, 8010 Graz, Austria}
\author{R. Casta\~{n}eda-Priego}
\affiliation{Instituto de F\'isica,
Universidad de Guanajuato, 37150 Leon, Mexico}
\date{\today}

\begin{abstract}
Our interest goes to the different virial contributions to the equation of state of 
charged colloidal suspensions.
Neglect of surface effects in the computation of the colloidal virial term leads to
spurious and paradoxical results. 
This pitfall is one of the several facets of the danger of a naive
implementation of the so called One Component Model, where the micro-ionic
degrees of freedom are integrated out to only keep in the description the
mesoscopic (colloidal) degrees of freedom.
On the other hand,
due incorporation of wall induced forces dissolves the paradox brought 
forth in the naive approach,
provides a consistent description, 
and confirms that for salt-free systems, the 
colloidal contribution to the pressure is dominated by the micro-ionic one. 
%This is at the root of the accuracy of simplified one colloid approached such as 
%cell or jellium-like models, compared to primitive model simulations or experiments. 
Much emphasis is put on the no salt case but the situation with added electrolyte 
is also discussed.
\end{abstract}
\maketitle

%%%%%%%%%%%%%%%%%%%%%%%%%%%%%%%%%%%%%%%%%%%%%%
\section{Introduction}
In a complex mixture where several species with widely different characteristic time
and space scales coexist, it is common practice to resort to a coarse grained description
integrating from the partition function all degrees of freedom that do not belong
to the main (larger) constituent \cite{Belloni,Hansen,Likos,Levin}. 
This leads to a state dependent effective
Hamiltonian for the main constituent, thereby allowing a One Component Model
(OCM) description. The motivation for such a procedure is not only to facilitate
contact with experiments, where most of the time the small constituents
cannot be probed directly, but also to simplify the theoretical treatment.
Indeed, one can then use  the well developed statistical mechanics tools from the
theory of simple liquids to study the OCM. This transposition 
from simple to complex fluids is however
paved with practical difficulties, see e.g. \cite{Louis,Baus}. It is the purpose of the
present paper to discuss one such pitfall arising in the context of charged 
colloidal suspensions.

The system we will consider is made up of 
$N_c$ charged spherical hard particles (colloids)
immersed in a solvent with dielectric constant $\varepsilon$, which fills
a box with volume $V$ limited by a neutral hard wall. 
The colloid's interior is assumed to have the same dielectric
constant as the solvent.  
Each colloid bears a charge $Z_c e$ where $e$ is the elementary charge
and $Z_c \gg 1$.
The medium outside
the container is a structureless dielectric continuum with dielectric constant
$\varepsilon'$. To ensure electroneutrality, 
the solution contains $N_c Z_c$ microscopic counterions, assumed monovalent.
Additional microions may also be present due to  the dissociation of an
added salt and the total number of microions is denoted 
$N_{\text{micro}}$. The particles interact through Coulomb forces and
hard sphere exclusion, although in the subsequent analysis,
the hard-core interaction will turn out to be irrelevant.

The paper is organized as follows. In the situation where 
$\varepsilon=\varepsilon'$, we consider in section \ref{sec:eos} the different virial 
contributions to the equation of state. In the salt-free case, 
we argue that the colloidal contribution $P_{\text{ocm}}$
has to be negligible compared to the microionic one ($P_{\text{micro}}$). 
We then show that a naive 
implementation of the OCM picture leads to a violation of the
constraint $P_{\text{ocm}} \ll P_{\text{micro}}$. 
Sections \ref{sec:nodiscont} for $\varepsilon = \varepsilon'$ and \ref{sec:disc} 
for $\varepsilon'<\varepsilon$ are devoted to the resolution of this 
apparent paradox. 
It will be shown that in a closed cell, the surface contribution to 
the colloidal 
virial $P_{\text{ocm}}$ is comparable to the bulk term, while only the latter is 
considered in the naive picture. Hence its failure, resulting from a
gross overestimation of $P_{\text{ocm}}$.
As a consequence, the knowledge of
a good effective potential in the bulk is insufficient when it comes 
to directly computing the colloidal virial in a closed box. 
Concluding remarks are drawn
in section \ref{sec:concl}, where we discuss in particular how the effective
potential can be used --indirectly but from a standard procedure-- 
to compute the total pressure of the system.
While most of the analysis holds without salt, the situation of an added electrolyte
is also briefly addressed.

%%%%%%%%%%%%%%%%%%%%%%%%%%%%%%%%%%%%%%%%%%%%%
\section{Equation of state, effective interactions and One Component Model virial}
\label{sec:eos}

\subsection{The equation of state}
We start by the simplest situation where $\varepsilon=\varepsilon'$ and consider
{\em all} charged species in the solution. The virial theorem allows to write the
total osmotic pressure $P$ (with respect to pure solvent) in the form
\be
\beta PV \,=\, N_c + N_{\text{micro}} \,+\, \frac{\beta}{3} \left\langle
\sum_{i \in col+micro} 
{\bm r}_i \cdot {\bm F}_i^{\text{int}} 
\right\rangle,
\label{eq:1}
\ee
where $\beta=1/(kT)$ is the inverse temperature and the summation runs 
over colloids and microscopic ions, therefore involving $N_c+N_{\text{micro}}$
terms. In Eq. (\ref{eq:1}) the angular brackets denote a statistical average 
(that coincides with time average) and ${\bm F}_i^{\text{int}}$ is the (internal) force
exerted on particle $i$ at position ${\bm r}_i$, due to hard core and Coulombic 
interactions in the solution. 
Adding the force exerted by the wall to 
${\bm F}_i^{\text{int}}$ would therefore provide the total force 
${\bm F}_i^{\text{tot}}$ felt by particle $i$.
We note here that it is possible to express the
pressure in Eq. (\ref{eq:1}) as a surface integral over the wall of the total
(colloid + micro-ions) concentration.
Applying the virial theorem to the microions only, we have
\be
N_{\text{micro}} kT + \frac{1}{3} \left\langle
\sum_{i \in micro} 
{\bm r}_i \cdot {\bm F}_i^{\text{tot}} 
\right\rangle
\,=\,0 \,=\,
N_{\text{micro}} kT + \frac{1}{3} \left\langle
\sum_{i \in micro} 
{\bm r}_i \cdot {\bm F}_i^{\text{int}} 
\right\rangle -
\frac{kT}{3} \left\langle\oint_{\text{box}}
\rho_{\text{micro}}({\bm r}) \, 
{\bm r}\!\cdot\! d^2{\bm S}\right\rangle,
\label{eq:2}
\ee
where the surface integral with
normal oriented outward runs over the box confining the system.
Inserting the latter equality into (\ref{eq:1}), we obtain
\begin{equation}
P \,=\, \rho_c \,kT + P_{\text{ocm}} + P_{\text{micro}}
\quad \hbox{with} \quad P_{\text{ocm}} = \frac{1}{3V}
\left\langle\sum_{i\in col} {\bm r}_i \!\cdot\! {\bm F}_i^{\text{int}} 
\right\rangle ~;
\quad P_{\text{micro}} = \frac{kT}{3V} \left\langle\oint_{\text{box}}
\rho_{\text{micro}}({\bm r}) \, 
{\bm r}\!\cdot\! d^2{\bm S}\right\rangle,
\label{eq:P}
\end{equation}
where $\rho_c=N_c/V$ and
$\rho_{\text{micro}}({\bm r})$ denotes the total microion
density at point ${\bm r}$.  
Within mean-field approximation, this equation may be found in \cite{Belloni}.
The first term on the right hand side 
of Eq. (\ref{eq:P}) is the colloid ideal gas term which
can be safely neglected in practice for the parameter range of
interest here (see below).  The second term --of central interest here-- is
the colloid-colloid virial contribution and is indexed by the
subscript OCM since it would be the only term considered (apart from
the ideal gas one) in the OCM approach, restricted to the
mesoscopic degrees of freedom $\{{\bm r}_i\}_{1\leq i\leq N_c} $.  
Indeed, the statistical average $\langle\ldots\rangle$ may be performed in two steps :
\be
P_{\text{ocm}} \,=\, \frac{1}{3V}
\left\langle\sum_{i=1}^{N_c} {\bm r}_i \!\cdot\! {\bm F}_i^{\text{int}} 
\right\rangle_{col+micro} \,=\, 
\frac{1}{3V}
\left\langle\sum_{i=1}^{N_c} {\bm r}_i \!\cdot\! \left\langle{\bm F}_i^{\text{int}}
\right\rangle_{micro} 
\right\rangle_{col}
\,=\, \frac{1}{3V}
\left\langle\sum_{i=1}^{N_c} {\bm r}_i \!\cdot\! {\bm F}_i^{\text{eff}} 
\right\rangle_{col},
\label{eq:Pocm}
\ee
where we have introduced the microion averaged effective
force ${\bm F}_i^{\text{eff}} $ exerted on colloid $i$ for a given colloid 
configuration.

The third term in
(\ref{eq:P}), $P_{\text{micro}}$, 
accounts for the direct coupling between colloids and microions. In
principle, this third term is to be averaged over the colloidal
degrees of freedom. However, even at simplified or 
mean-field level, a full $N_c$-colloid
simulation is computationally demanding
\cite{Fushiki,LHM,JH}, and further simplifications are helpful.
Of particular interest are two such simplifications, both belonging to
the Poisson-Boltzmann family,
that reduce the initial
$N_c$-body problem onto a $N_c=1$ body situation. The first one is 
the common cell model approach originating from a
solid state point of view where the Wigner-Seitz cell around a colloid
is constructed and then ``sphericalized'' for the sake of
simplicity. The Poisson-Boltzmann equation is solved within this cell, and from the
microionic density profile one can then estimate $P_{\text{micro}}$. 
The second model is the renormalized jellium model \cite{Jellium} where a
liquid state point of view is adopted: the colloid-colloid pair
distribution function $g_{cc}(r)$ is considered structureless so that
other colloids around a tagged macroion behave as a continuous
background. The charge of this background is {\it a priori} unknown,
and enforced to coincide with the effective charge. This
self-consistency requirement leads to a unique and well defined
effective charge \cite{Jellium}.  It has been shown that for salt-free
suspensions, these two models -- cell and jellium -- both lead to a
pressure $P_{\text{micro}}$ that is in excellent agreement with
existing experimental data \cite{Reus} and primitive model simulations
for $P$ \cite{Linse,rque4}, see e.g. \cite{LevinJPCM,Jellium,nota}.  We note
that $P_{\text{micro}}$ may be coined a ``volume'' term
\cite{volume,Rene,Hansen}, since -- at least, within the cell model and
jellium approaches -- it does not depend on the colloidal degrees of
freedom but only on the mean colloidal density.
The good agreement one obtains with the exact pressure $P$ for both
models implies that for salt-free systems $P\simeq P_{\text{micro}}$. 
This is corroborated by a recent study of finite stiff-chain 
polyelectrolytes \cite{Antypov}. From 
Eq. (\ref{eq:P}) where the ideal gas contribution ($\rho_c kT$) is neglected, this may 
be transposed into the following requirement: 
\be
P_{\text{ocm}} \ll P_{\text{micro}}.
\label{eq:constraint}
\ee 
A similar conclusion was reached in Ref. \cite{ecc}.

\subsection{Effective interactions}
Both Poisson-Boltzmann cell and jellium approaches are not only useful to estimate the pressure,
but also to derive effective parameters for solvent + microions
averaged colloid/colloid interactions.  By construction, the effective
potential is that which leads to the correct colloid-colloid pair
structure encoded in the potential of mean force
$g_{cc}$, {\em assuming pair-wise colloid-colloid
interactions within the OCM model} (see e.g. \cite{Belloni}).  
Although the effective
potential has a clear-cut definition, there is no rigorous {\em
operational} route to construct this object.
In general, when microionic correlations do not
invalidate the mean-field picture \cite{rque4},
 a good approximation is to write
the effective potential as a sum of pair-wise Yukawa terms of the form
\begin{equation}
\beta \,v_{\text{eff}}(r)\,=\, \Zeff^2  \lambda_B 
\left(\frac{\exp(\kapeff a)}{1+\kapeff a}\right)^2
\frac{\exp(-\kapeff r)}{r}
\label{eq:DLVO}
\end{equation}
with $a$ the colloid radius,
$\lambda_B = \beta e^2/\varepsilon$ the Bjerrum length, and $\Zeff$ and $\kapeff$ the
effective charge and inverse screening length computed within the cell
or jellium model \cite{Alexander,bell00,Lang03,Jellium}.  
Such a ``DLVO''-like expression \cite{Belloni,Hansen,Levin} would accurately
reproduce the large distance interaction of two colloids in a salt sea
\cite{Belloni,Levin,Hansen}.
Its relevance in the no-salt case will not be discussed. 
As will become clear below, we are interested here in orders
of magnitude, that should not depend on the precise form of 
(\ref{eq:DLVO}).

%%%%%%%%%%%%%%%%%%%%%%
\subsection{An apparent paradox}
Within the jellium model, the salt-free 
equation of state takes a particularly simple form
\be
\beta P_{\text{micro}}= \Zeff \rho_c.
\label{eq:300} 
\ee
Within the cell model, this
expression is not exact but approximately correct. For a highly
charged macroion, one has $\Zeff \gg 1$ which allows to neglect the
ideal gas term in (\ref{eq:P}).  In spite of its simplicity, the
expression $\beta P_{\text{micro}} = \rho_c\,\Zeff $ hides a complex
density dependence through $\Zeff$ and is in excellent agreement with
the exact pressure $P$ found experimentally or in primitive model
simulations, as emphasized above. In addition, the effective screening
length reads \cite{Jellium}
\be
\kapeff^2 = 4 \pi \lambda_B \,\rho_c\,\Zeff
\label{eq:kappa}
\ee
The constraint embodied in Eq. (\ref{eq:constraint})
may therefore be rewritten
\be
\beta P_{\text{ocm}} \ll \Zeff \rho_c.
\label{eq:constraintbis}
\ee
Alternatively, in the low electrostatic coupling regime (where $\Zeff$ coincides
with $Z_c$), one should recover the ideal gas pressure 
$\beta P \simeq \rho_c (1+\Zeff)$. Given that in this limit, 
$\beta P_{\text{micro}} \simeq \rho_c \Zeff$, we recover the requirement 
(\ref{eq:constraintbis}), that will be an important benchmark for the
following analysis. We now turn to the formulation of the apparent paradox.

In  the bulk of the suspension, the effective potential (\ref{eq:DLVO}) provides the effective force
acting on a colloid $i$
\be
{\bm F}_i^{\text{eff}} \,=\, \sum_{j=1}^{N_c} {\bm F}_{ij}^{\text{eff}} \,=\,
-\sum_{j=1}^{N_c} \,
{\bm \nabla}_{\bm r} \, v_{\text{eff}}(r) \biggl|_{{\bm r}={\bm r}_i-{\bm r}_j}.
\label{eq:Fpaire}
\ee
Considering naively that $P_{\text{ocm}}$
appearing in (\ref{eq:P}) and (\ref{eq:Pocm})
is dominated in a very large system by its bulk behaviour,
we insert (\ref{eq:Fpaire}) into (\ref{eq:Pocm}) to approximate
$P_{\text{ocm}}$ by $P_{\text{ocm}}^*$ with
\be
P_{\text{ocm}}^*  \,=\,  \frac{1}{3V}
\left\langle\sum_{i,j=1}^{N_c} {\bm r}_{i} \!\cdot\! {\bm F}_{ij}^{\text{eff}} 
\right\rangle_{col}
\,=\,  \frac{1}{6V}
\left\langle\sum_{i,j=1}^{N_c} {\bm r}_{ij} \!\cdot\! {\bm F}_{ij}^{\text{eff}} 
\right\rangle_{col},
\ee
where ${\bm r}_{ij}={\bm r}_i-{\bm r}_j$. We will subsequently omit
the subscript ``col'' indicating the degrees of freedom involved in the average.
Introducing the colloid-colloid pair correlation function $g_{cc}(r)$,
we can write
\begin{eqnarray}
\beta P_{\text{ocm}}^* &=& -\frac{\rho_c^2}{6} \,\int_{r=2a}^\infty
g_{cc}(r)\, \frac{d\beta v_{\text{eff}}(r)}{dr}\,r\,d^3{\bm
r}
\\
&=& \frac{2 \pi\,\rho_c^2 \,\Zeff^2 \,\lambda_B}{\kapeff^2} \left\{
1+\frac{(\kapeff a)^2}{3(1+\kapeff a)^2}\right\}
+ 
\frac{\rho_c^2 }{6} \int_{r=2a}^\infty
[g_{cc}(r)-1] \,(1+\kapeff r) \, \beta  v_{\text{eff}}(r)\,d^3{\bm
r}. 
\label{eq:ocmvirial2}
\end{eqnarray}
To estimate the above quantity, it is sufficient to keep the dominant term
only, which is the first one on the rhs, arising from the long-range
behavior of the pair correlation function ($g_{cc}\to 1$ at large
distances). In this term, the curly brackets may be safely
approximated by 1 since at low densities, $\kapeff a \ll 1$.  
%Such approximations are fully justified in the dilute limit.
Remembering Eq. (\ref{eq:kappa}), we obtain
\begin{eqnarray}
P_{\text{ocm}}^*  & \simeq &
\frac{2 \pi\,\rho_c^2 \,\Zeff^2 \,\lambda_B}{\kapeff^2}
\label{eq:Pstar1}
\\
&\simeq &
\frac{1}{2} \Zeff \rho_c,
\label{eq:Pstar2}
\end{eqnarray}
The factor $1/2$ which appears is classical (see e.g. \cite{Belloni}).
The important point here is that estimation (\ref{eq:Pstar2})
by far violates the constraint (\ref{eq:constraintbis}). 
A similar conclusion would be reached 
including the first correction in $\Zeff^2 \exp(-\kapeff r)/r$ 
to the long distance behaviour
$g=1$ when computing the integral on the rhs of (\ref{eq:ocmvirial2}):
this yields $P_{\text{ocm}}^* \simeq \Zeff \rho_c/2[1 + {\cal O}(\kapeff \lambda_B)]$
with $\kapeff \lambda_B \ll 1$ in the dilute limit. 
The paradox here is that the very same approach that provides
a contribution $P_{\text{micro}}$ very close to the total pressure,
gives an effective potential that {\em apparently} spoils the previous agreement,
by grossly overestimating the colloidal virial contribution to the pressure.
We will see that this feature is not ascribable to a failure
of the functional form of Eq. (\ref{eq:DLVO}),
which provides a decent approximation for the quantity 
$P_{\text{ocm}}^*$.

%%%%%%%%%%%%%%%%%%%%%%%%
\subsection{How can the paradox be resolved ?}
The root of the paradox reported above is that
approximating $P_{\text{ocm}}$ by $P_{\text{ocm}}^*$ is 
{\em incorrect}: while $P_{\text{ocm}}^*$ provides a reasonable
estimate for the bulk contribution to $P_{\text{ocm}}$, {\em surface effects}
make that in the vicinity of the wall, the effective force felt by a colloid
differs from (\ref{eq:Fpaire}). These surface 
induced terms play a key role here and contribute a large
amount to the colloidal virial $P_{\text{ocm}}$, {\em no matter how large
the system is}. It turns out that bulk and surface induced contributions
almost cancel each other, so that  the resulting expression for
$P_{\text{ocm}}$ is much smaller than $P_{\text{ocm}}^*$ and therefore
fulfills the requirement (\ref{eq:constraintbis}).
Our goal in the remainder is to illustrate this cancellation explicitly,
from a correct description of confinement effects.
To this aim, it is judicious to simplify the problem by 
considering the limit of point
colloids ($a=0$), and by identifying the effective charge  
with the bare one $Z_c$. Considering charge renormalization effects
is here immaterial and focussing on dilute systems where $\kappa a$ is
small, finite $a$ effects do not affect our main conclusions.
In the bulk of the suspension,
the effective potential therefore takes a simple Yukawa form
\begin{equation}
\beta \,v_{\text{eff}}(r)\,=\, Z_c^2  \lambda_B 
\frac{\exp(-\kappa r)}{r},
\end{equation}
with $\kappa^2=4 \pi \lambda_B Z_c \rho_c$.

At this point, a comparison with simple electrolytes seems appropriate,
for the aforementioned cancellation is already present.
For our discussion, we may consider that the role of the colloids
is played by the cations, and that the anions constitute the 
remaining ``microions''. The pressure 
has to be close to \cite{Belloni,Levin}
\be
\beta P_{\text{electrolyte}} \simeq \rho_{\text{anion}} + \rho_{\text{cation}} -\frac{\kappa^3}{24 \pi},
\label{eq:15}
\ee
with equal mean densities $\rho_{\text{anion}}=\rho_{\text{cation}}$. From the contact theorem,
we deduce the densities at the wall
\be
\rho_{\text{anion}}(wall) = \rho_{\text{cation}}(wall) \simeq 
\rho_{\text{anion}}  -\frac{\kappa^3}{48 \pi}.
\label{eq:16}
\ee
Rewriting (\ref{eq:P}) in the form
\begin{equation}
\beta P_{\text{electrolyte}} \,=\, \rho_{\text{cations}} +
 \frac{\beta}{3V}
\left\langle\sum_{i\in cation} {\bm r}_i \!\cdot\! {\bm F}_i^{\text{int}} 
\right\rangle
+ \rho_{\text{anion}}(wall)
\end{equation}
we obtain from (\ref{eq:15}) and (\ref{eq:16})
\begin{equation}
 \frac{\beta}{3V}
\left\langle\sum_{i\in cation} {\bm r}_i \!\cdot\! {\bm F}_i^{\text{int}} 
\right\rangle\,\simeq\,  -\frac{\kappa^3}{48 \pi}.
\label{eq:18}
\end{equation}
Given that $\kappa^2 = 8 \pi \lambda_B Z^2 \rho_{\text{cation}}$,
we have $\kappa^3/(\beta P_{\text{ocm}}^*) \propto \kappa \lambda_B $
which is a small quantity for a dilute system.
We explicitly see here that the ``colloidal'' virial [lhs of (\ref{eq:18}) up to a factor 
$\beta$] 
is by far smaller than 
the estimation $P_{\text{ocm}}^*$.

%%%%%%%%%%%%%%%%%%%%%%%%%%%%%%%%%%%%%%%%%%%%%
\section{Wall mediated forces without dielectric discontinuity}
\label{sec:nodiscont}

In the vicinity of the wall, the colloids do not see a spherically symmetric
environment. As a consequence,
\begin{enumerate}
\itemsep=0pt
\item the usual $\exp(-\kappa r)/r$ pair interaction is modified.
\item the mean force acting on a colloid does not vanish. This is a one body,
wall induced effect, mediated by the microions. It is therefore an 
{\em internal} force, that should be taken into account in
(\ref{eq:Pocm}). It should not be
confused with the {\em external} (and short range) direct colloid-wall interaction.
\end{enumerate}
Evaluating the rhs of (\ref{eq:Pocm}) therefore 
requires a careful computation of
both types of microion averaged colloidal forces.
To this end, we need the solution $\phi_z(\rho,z')$
of Debye-H\"uckel
equation $\nabla^2 \phi_z = \kappa^2  \theta(z') \phi_z$ in the case where a test charge
is located in the solution a distance $z$ from an infinite neutral wall.
We have introduced the Heaviside function $\theta$ and cylindrical coordinates ($\rho,z'$) such that
the test particle is located at ($0,z$) with $z>0$. 
The planar geometry approximation for the wall is sufficient provided the cell size
or radius of curvature is much larger than Debye length $1/\kappa$.
We start by the situation of equal dielectric constants inside
and outside  the solution ($\varepsilon=\varepsilon'$). 
The electrostatic potential may be written in the form of a Hankel 
(two dimensional Fourier) transform
\cite{Bracewell}
where $q$ and $\rho$ are conjugate quantities
\cite{Janco,Netz99}
\be
\phi_z(\rho,z')\,=\, Z_c \lambda_B \int_0^\infty \left(
\frac{k-q}{k+q}\,e^{-k(z+z')} \,+\,e^{-k |z-z'|}
\right)\frac{1}{k}\,J_0(q\rho) \, q \,dq
 \quad ;\quad k \equiv \sqrt{\kappa^2+q^2}
 \label{eq:pot}
\ee
The second term in the integrand ($e^{-\kappa|z-z'|}$) gives exactly
$Z_c \lambda_B \exp(-\kappa r)/r$ where $r=[\rho^2+(z-z')^2]^{1/2}$ is the distance to the
source. This is the standard Debye-H\"uckel potential which dominates
in the bulk. The remaining term, which vanishes at large distances
($\kappa z$ or $\kappa z' \gg 1$) is due to the presence of the interface.

%%%%%%%%%%%%%%%%%%%%
\subsection{One colloid ion average force}
The force felt by a colloid located a distance $z$ from the planar interface
follows from (\ref{eq:pot}), considering the electrostatic potential 
$\widetilde\phi_z=\phi_z-Z_c^2\lambda_B \,e^{-\kappa r}/r$ where the self term
has been subtracted:
\be
\beta {\bm F}_{c-wall} \,=\,{\mathbf{ \widehat n}} \, Z_c \frac{\partial}{\partial z'}
 \beta \widetilde \phi_z(0,z')\biggl|_{z,z'=z}
\,=\, Z_c^2 \, \lambda_B \, \int_0^\infty 
\frac{k-q}{k+q}\,e^{-2 k z} 
 \, q \,dq \,(-{\mathbf{ \widehat n}}).
\label{eq:f1}
\ee
In this equation, ${\mathbf{ \widehat n}}$ denotes the unit vector
perpendicular to the interface pointing outside the solution.
We coin the force (\ref{eq:f1})  ``colloid-wall'' and for notational convenience,
we henceforth omit the superscripts ``int'' and ``eff''. 
This force repels the colloid
from  the wall [$k=(\kappa^2+q^2)^{1/2}>q$], 
as a result of microions imbalance between the half
of the colloid exposed to the wall, and the other hemisphere. 
Inserting (\ref{eq:f1})
into (\ref{eq:Pocm}) we have
\be
\frac{1}{3V}\, \left\langle\sum_{i=1}^{N_c} {\bm r}_{i} \!\cdot\! {\bm F}_{i-wall} 
\right\rangle = \frac{1}{3V} \int_{wall}\! d^2S \,\int_0^\infty \rho_c(z) \,
{\bm r}\!\cdot\! {\bm F}_{i-wall}(z)\, dz.
\label{eq:int1}
\ee
To leading order, the above integral may be computed assuming a
uniform density of colloids $\rho_c(z)=\rho_c$. In (\ref{eq:int1}),
$\bm r$ denotes the absolute position with ${\bm r} = {\bm s}- z \,\mathbf{\widehat n}$
(${\bm s}$ is therefore the orthogonal projection of $\bm r$ onto the wall). We neglect
the term in $- z \,\mathbf{\widehat n}$ (that would contribute proportionally
to the surface of the system), so that 
\begin{eqnarray}
\frac{1}{3V}\, \left\langle\sum_{i=1}^{N_c} {\bm r}_{i} \!\cdot\! \beta{\bm F}_{i-wall} 
\right\rangle &\simeq& \frac{\beta \rho_c}{3V} \int_{wall}\! d^2S \,\int_0^\infty  
{\bm s}\!\cdot\! {\bm F}_{i-wall}(z)\,dz 
\label{eq:22}\\
&\simeq&- \frac{\rho_c}{3V} \, Z_c^2 \lambda_B 
\left(\int_{wall} {\bm s}\cdot \mathbf{\widehat n} \,d^2 S\right) \,
\int_0^\infty  dz \int_0^\infty 
\frac{\sqrt{\kappa^2+q^2}-q}{\sqrt{\kappa^2+q^2}+q}\,e^{-2 z \sqrt{\kappa^2+q^2} } 
 \, q \,dq \\
 &\simeq& - \frac{1}{6} \,\rho_c Z_c^2 \,\kappa \lambda_B\\
 &\simeq& -\frac{\kappa^3}{24 \pi} .
 \label{eq:25}
\end{eqnarray}
Incidentally, this is exactly the Debye-H\"uckel form for the excess pressure
of an electrolyte 
[see Eq. (\ref{eq:15})]. For dilute systems, this quantity is small compared
to $\rho_c Z_c$, as emphasized earlier. The constraint
(\ref{eq:constraintbis}) is therefore fulfilled.

%%%%%%%%%%%%%%%%%%%%
\subsection{Colloid-colloid interactions}
\label{ssec:colcol}

Within the simple Debye-H\"uckel treatment, the potential of interaction
between two colloids near the wall (one at $z$, the other at $z'$, with a lateral
distance $\rho$ between them) is $Z_c \phi_z(\rho,z') = Z_c\phi_{z'}(\rho,z)$. 
To calculate the force felt by the colloid at $z$ due to all neighbors,
we assume again a uniform distribution of neighbors:
\begin{eqnarray}
\beta {\bm F}_{col-col}(z) &=& {\mathbf{ \widehat n}} \,\rho_c \int_{z'=0}^\infty dz'\,
\int_0^\infty 2 \pi \rho \,d\rho\,
 \frac{\partial  Z_c \beta  \phi_z(\rho,z')}{\partial z}\biggl|_{z'} \\
 &=&  -{\mathbf{ \widehat n}} \,\rho_c Z_c^2 \lambda_B \int_{0}^\infty dq\,
\int_0^\infty 2 \pi \rho \,d\rho\, 
\left(\frac{\sqrt{\kappa^2+q^2}-q}{\sqrt{\kappa^2+q^2}+q}-1 \right)\,e^{-z\sqrt{\kappa^2+q^2}  } \,\frac{1}{\sqrt{\kappa^2+q^2}} \,J_0(q \rho)
 \, q \,dq
 \label{eq:fcc1}
\end{eqnarray}
The component of the force parallel to the wall vanishes upon averaging.

Inserting this force into (\ref{eq:Pocm}) and proceeding along similar lines as in
Eqs. (\ref{eq:22}) sq, we have 
\be
\frac{\beta}{3V}\, \left\langle\sum_{i=1}^{N_c} {\bm r}_{i} \!\cdot\! {\bm F}_{i-coll} 
\right\rangle \,\simeq \, \rho_c^2 Z_c^2 \lambda_B 
\int_0^\infty 2 \pi \rho\,d\rho\, \int_0^\infty 
\left(-\frac{\sqrt{\kappa^2+q^2}-q}{\sqrt{\kappa^2+q^2}+q}+1 \right)
 \,\frac{1}{\kappa^2+q^2} \,J_0(q \rho)
 \, q \,dq.
 \label{eq:hankel2}
\ee
Both expressions (\ref{eq:fcc1}) and (\ref{eq:hankel2}) are of the form 
of a Hankel transform at the origin $q=0$ of the inverse Hankel transform
of a function $A(q)$, with $A=(\ldots-1)e^{-k z}/k$ in (\ref{eq:fcc1})
and $A = (-\ldots+1)/k^2$ in (\ref{eq:hankel2}). This is nothing
but $A(0)$ \cite{Bracewell}, which vanishes in both cases. 
Therefore, with the approximations proposed, the force
in (\ref{eq:fcc1}) and the virial term in (\ref{eq:hankel2}) vanish.
To be more specific, we compute explicitly the integrals in (\ref{eq:hankel2}):
\be
\frac{\beta}{3V}\, \left\langle\sum_{i=1}^{N_c} {\bm r}_{i} \!\cdot\! {\bm F}_{i-coll} 
\right\rangle \,\simeq \,\frac{1}{2}\, \rho_c Z_c \, (-1 +1).
\label{eq:fcc2}
\ee
The term in +1 in the parenthesis arises from the term in +1 in Eq. (\ref{eq:hankel2}),
which gives the usual ``bulk'' $e^{-\kappa r}/r$ pair interaction,
as already mentioned. The associated
virial is $\beta P_{\text{ocm}}^*= Z_c \rho_c /2$, as obtained 
in (\ref{eq:Pstar2}). The present calculation shows that this term is canceled
by an opposite wall induced contribution. If the simplifying assumption 
$g_{cc}=1$  is relaxed, the resulting expression for (\ref{eq:fcc2})
no longer vanishes but remains negligible with respect to $\rho_c Z_c$.
On the other hand, relaxing the assumption of a uniform profile
$\rho_c(z)$ leaves the result unaffected, as will be seen in section
\ref{sec:disc}.

We conclude here that summing the two contributions from Eqs.
(\ref{eq:25}) and (\ref{eq:fcc2}) provides a value for 
$P_{\text{ocm}}$ that is compatible with the constraint
(\ref{eq:constraintbis}).

%%%%%%%%%%%%%%%%%%%%%%%%%%%%%%%%%%%%%%%%%%%%
\section{Analysis in presence of a dielectric discontinuity}
\label{sec:disc}

In this section, we extend the previous analysis to the situation where
the dielectric constants are not matched: $\eta = \varepsilon' / \varepsilon \neq 1$.
The relevant parameter range corresponds to $\eta<1$
e.g for  water droplets in air in a spray-drying             
experiments. The first important difference
with the $\eta=1$ case is that the equation of state (\ref{eq:P}) takes
a different form. The pressure is indeed not solely given by the contact 
densities of charged species at the wall, but contains additional 
electric contributions (polarization or image effects). On the other
hand, Eqs (\ref{eq:1}) and (\ref{eq:2}) are still formally correct 
provided one also includes in the ``internal'' forces the electric forces from the wall. 
The resulting equation of state reads
\begin{equation}
P \,=\, \rho_c \,kT+ \frac{1}{3V}
\left\langle\sum_{i\in col} {\bm r}_i \!\cdot\! {\bm F}_i^{\text{int}} 
\right\rangle + \frac{kT}{3V} \left\langle\oint_{\text{box}}
\rho_{\text{micro}}({\bm r}) \, 
{\bm r}\!\cdot\! d^2{\bm S}\right\rangle
+ \frac{1}{3V}\left\langle
\oint_{\text{box}} {\bm r} \!\cdot {\bm T}^{el}\, d{\bm S}
\right\rangle.
\label{eq:Peta}
\end{equation}
Here 
\be
{\bm T}^{el} = \frac{\varepsilon}{8\pi}\ E^2 {\bm I} - \frac{\varepsilon}{4\pi}\,
{\bm E}\otimes{\bm E}
\ee
is the Maxwell tensor, with ${\bm E}$ the local electric field and ${\bm I}$ 
the isotropic tensor.

The counterpart of (\ref{eq:pot}) now reads:
\be
\phi_z(\rho,z')\,=\, Z_c \lambda_B \int_0^\infty \left(
\frac{k-\eta q}{k+\eta q}\,e^{-k(z+z')} \,+\,e^{-k |z-z'|}
\right)\frac{1}{k}\,J_0(q\rho) \, q \,dq
 \quad ;\quad k \equiv \sqrt{\kappa^2+q^2}.
 \label{eq:poteta}
\ee
As in the case $\eta=1$ and as long as $\eta\neq 0$, 
the corresponding interaction between two colloids decays as $\rho^{-3}$  
at large distances parallel to the wall (see \cite{Wurger} for a discussion 
of this dipolar-like term).
When $\eta=0$, the wall can be formally removed considering the
electric image located symmetric to the $z=0$ plane.

The colloid-colloid and colloid-wall interactions readily follow from 
(\ref{eq:poteta}). At short distances $z\to 0$, the latter diverges like
$z^{-1} (1-\eta)/(1+\eta)$ \cite{Netz99}, which corresponds to the 
unscreened interaction of a particle with its own image. 
This divergence means that the uniform colloid density cannot be invoked 
when it comes to computing (\ref{eq:int1}). To obtain the leading order behaviour,
we can assume that the colloids are distributed with the Boltzmann
weight $\rho_c (z)=\rho_c\exp[-\beta \phi_{c-wall}(z) ]$,
where ${\bm F}_{c-wall}=-{\bm \nabla} \phi_{c-wall}$ and the potential
$\phi_{c-wall}$ deriving from (\ref{eq:poteta}) vanishes for $z\to \infty$.
The precise knowledge of this potential is however not required since
\begin{eqnarray}
\frac{1}{3V}\, \left\langle\sum_{i=1}^{N_c} {\bm r}_{i} \!\cdot\! {\bm F}_{i-wall} 
\right\rangle &=& \frac{\rho_c}{3V} \int_{wall}\! d^2S \,\int_0^\infty  \,
{\bm r}\!\cdot\! {\bm F}_{c-wall}(z)\,\exp[-\beta \phi_{c-wall}(z)] \,dz.
\\
&\simeq& \rho_c \, kT\, \left[\exp(-\beta \phi_{c-wall}(z))
\right]_0^\infty \\
&\simeq& -\rho_c kT.
\end{eqnarray}
This term therefore cancels the ideal gas one on the rhs of (\ref{eq:Peta}).

The wall induced colloid-colloid contribution to the colloidal virial
may be computed along similar lines as in section \ref{ssec:colcol}.
An expression involving again a Hankel transform composed with its inverse
is again obtained, with now a function 
\be
A(q)=\left(\frac{\sqrt{\kappa^2+q^2}- \eta q}{\sqrt{\kappa^2+q^2}+ \eta q}-1 \right)
\,\frac{1}{\sqrt{\kappa^2+q^2}}
\int_0^\infty dz\, \rho_c(z)
e^{-z\sqrt{\kappa^2+q^2}  } 
\ee
Since $A(0)=0$, we conclude here that 
\be
\left\langle\sum_{i=1}^{N_c} {\bm r}_{i} \!\cdot\! {\bm F}_{i-coll} 
\right\rangle \,\simeq \, 0,
\ee
so that the total colloidal virial [including colloid-colloid and colloid-wall
interactions] is close to $-\rho_c kT$, which is a small quantity
compared to the microionic contribution $Z_c \rho_c kT$. 
Equation (\ref{eq:Peta}) can finally be rewritten
\be
P \,\simeq\, \frac{kT}{3V} \left\langle\oint_{\text{box}}
\rho_{\text{micro}}({\bm r}) \, 
{\bm r}\!\cdot\! d^2{\bm S}\right\rangle
+ \frac{1}{3V}\left\langle
\oint_{\text{box}} {\bm r} \!\cdot {\bm T}^{el}\, d{\bm S}
\right\rangle.
\ee

%%%%%%%%%%%%%%%%%%%%%%%%%%%%%%%%%%%%%%%%%%%%
\section{Concluding remarks}
\label{sec:concl}

Before briefly discussing the situation 
where a salt is added, two comments are in order.

%%%%%%%%%%%%%%%%%%%
\subsection{Closed cells versus periodic boundary conditions}

From the previous discussion, it appears that the equation of state 
 (\ref{eq:P})
holds when the system is confined
by a hard wall, and would fail if periodic boundary conditions (pbc)
would be enforced. The inadequacy of $P_{\text{ocm}}^*$ to approximate
$P_{\text{ocm}}$ may then be phrased in the following way
\begin{eqnarray}
3 V P_{\text{ocm}}^* &\equiv&  \frac{1}{2}
\left\langle\sum_{i,j=1}^{N_c} \,\sum_{\mathbf{n}} \,
{\bm r}_{ij} \cdot
{\bm F}_{ij}^{\text{eff}} ({\bm r}_{ij}-{\bm R}_{\mathbf{n}})
\right\rangle_{pbc}
\label{eq:pbc1} \\
&\neq& 
\left\langle\sum_{i}^{N_c}  
{\bm r}_{i} \!\cdot\! {\bm F}_{i}^{\text{eff}} 
\right\rangle_{hard ~walls} 
\label{eq:pbc2}
\end{eqnarray}
where in (\ref{eq:pbc1}), the sum involves all periodic images
of the cell considered: $\mathbf{n}$ is a vector with components
in $\mathbb{Z}^3$,  which indexes the 
center $\mathbf{R}_{\mathbf{n}}$ of a given image of the ``central'' cell.
The central cell has $\mathbf{R}_{\mathbf{0}}=\mathbf{0}$ 
and since we deal here with a short
range effective potential, the sum over $\mathbf{n}$ may be truncated
to retain only the 7 terms with $|\mathbf{n}|\leq 1$.

However, for any simple fluid where the forces 
${\bm F}_{i}$ are given, (\ref{eq:pbc2}) would be an equality.
%, resulting from two equivalent expressions of the excess pressure (one from the 
%virial theorem, the other from the standard volume derivative of the free energy).
Indeed we have
\be
\left\langle\sum_{i} {\bm r}_{i} \!\cdot\! {\bm F}_{i}
\right\rangle_{hard ~walls}^{simple~fluid}  \,\equiv\,
\frac{1}{2}
\left\langle\sum_{i,j} {\bm r}_{ij} \!\cdot\! {\bm F}_{ij}
\right\rangle_{hard ~walls}^{simple~fluid}
\ee
where the rhs shows negligible dependence on the boundary conditions
provided the system is large enough, and can then be computed
with pbc provided the correct forces are considered
$[{\bm F}_i=\sum_j \sum_{\mathbf{n}} {\bm F}_{ij}({\bm r}_{ij}-{\bm R}_{\mathbf{n}})]$.
Hence
\be
\left\langle\sum_{i} {\bm r}_{i} \!\cdot\! {\bm F}_{i}
\right\rangle_{hard ~walls}^{simple~fluid}  \, = \,
\frac{1}{2}
\left\langle
\sum_{i,j} \,\sum_{\mathbf{n}} \,
{\bm r}_{ij} \cdot
{\bm F}_{ij} ({\bm r}_{ij}-{\bm R}_{\mathbf{n}})
\right\rangle_{pbc}^{simple~fluid}
\label{eq:pbc3}
\ee
The difference between equations (\ref{eq:pbc2}) and (\ref{eq:pbc3})
illustrates the important role of microions. We  may also consider that the
$\neq$ sign in (\ref{eq:pbc2}) arises from the density dependence of
the effective pair potential. 

A natural question at this point is : does the knowledge of the ``bulk'' effective potential
(\ref{eq:DLVO}) between colloids allow to compute their
virial $P_{\text{ocm}}$ as it appears in (\ref{eq:P}) ?
The answer is positive in a closed cell, at the OCM level, provided  that due account is taken for the dielectric images of the colloids. In the following section, we address
a related question, and discuss how the full pressure of the colloidal system
may be recovered, 
%from pbc simulations, 
assuming again that the only information at hand
is that of the bulk effective colloid-colloid interaction.

%%%%%%%%%%%%%%%%%%%
\subsection{Back to the DLVO potential}
\label{ssec:back}
We consider here a simple liquid that interacts with a pair-wise potential
given by Eq. (\ref{eq:DLVO}), with effective parameters 
$\Zeff^* \gg 1$ and $\kapeff^{*2} = 4 \pi \lambda_B \Zeff^* \rho_c^*$
(salt-free case, for simplicity). These parameters are fixed {\it a priori},
and chosen to coincide with those relevant for a colloidal suspension 
at $\rho_c=\rho_c^*$. The potential of interaction is therefore {\em density
independent} and the system, later referred to as ``auxiliary'',  
can be studied for
$\rho_c\neq \rho_c^*$. 

We consider the parameter range (essentially low density) where 
the excess pressure of such a system is well approximated
by $P_{\text{ocm}}^*$ in Eq. (\ref{eq:Pstar1}):
\be
\beta P_{\text{ocm}}^*  \, \simeq \,
\frac{2 \pi\,\rho_c^2 \,\Zeff^{*2} \,\lambda_B}{\kapeff^{*2}}
\,=\,
\frac{1}{2}\, \frac{\rho_c^{2}}{\rho_c^*} \,\Zeff^*.
\label{eq:Pstar3}
\ee 
Incidentally, 
the contact theorem indicates that the contact density in the case where 
the system is confined by a closed box, reads
$\rho_c(wall) \simeq \rho_c^2 \Zeff^* /(2 \rho_c^*)$. 
This quantity is much larger than the mean density $\rho_c$
(except when $\rho_c$ is extremely small, a limit of little interest here).
This excess with respect to the
mean density is to be contrasted with the {\em depletion} from the wall
that is present in the original colloidal system containing microions:
Eq. (\ref{eq:f1}) for $\varepsilon=\varepsilon'$ shows a repulsive colloid-wall
behaviour, and the depletion is even stronger when $\varepsilon'<\varepsilon$
due to like-sign images, see the discussion after Eq. (\ref{eq:poteta}).

The pressure of the simple liquid with DLVO interactions, close
to $P_{\text{ocm}}^*$, has {\it a priori} nothing to do with the pressure
$P^{\text{original}}$
of the real colloidal system. It has also nothing to do with the colloid
virial contribution entering Eq. (\ref{eq:P}). However, for $\rho_c=\rho_c^*$,
the colloid-colloid structural information is the same for both
original and auxiliary systems. One may then invoke 
Kirkwood-Buff identity \cite{KB} which states
that the inverse compressibility of the original colloidal suspension coincides with the
long wave-length limit of the colloid-colloid structure factor
$S_{cc}(k)$~:
\begin{equation} \chi
\,=\, \left(\frac{\partial \beta P^{\text{original}}}{\partial \rho_c}
\biggl|_T\right)^{-1}\,=\, S_{cc}(0).
\label{eq:kb}
\end{equation} 
The compressibility in our auxiliary simple liquid with {\em fixed} potential
of interaction is therefore the same at $\rho_c=\rho_c^*$
(and only at this density)
\be
\frac{\partial P^{\text{original}}}{\partial \rho_c} \biggl|_T 
~~  \,\stackrel{ \rho_c=\rho_c^*}{=}\, ~~
\frac{\partial P^*_{\text{ocm}}}{\partial \rho_c} \biggl|_{T,\kapeff^*,\Zeff^*}
\label{eq:exe}
\ee
This offers a means to compute the
equation of state of the original colloidal system from integrating the
inverse compressibility of the auxiliary one. In this integration, due account
must be taken of the density dependence of both $\Zeff^*$ and $\kapeff^*$.
The previous integration procedure therefore requires to consider the auxiliary 
system for several
values of $\rho_c$ for a given $\rho_c^*$ [to compute the derivative in the rhs
of (\ref{eq:exe})], before scanning the range
of interest for $\rho_c^*$. 
Of course, the general procedure outlined here does not depend on the
specific form of the effective potential, and is equally valid
when salt is added. It turns however that the DLVO
potential together with the salt-free approximation (\ref{eq:Pstar1})
--which leads to (\ref{eq:Pstar3})--
provide a clear illustration of the procedure.
From (\ref{eq:Pstar3}), we obtain the rhs of Eq. (\ref{eq:exe}):
\be
\frac{\partial\beta  P^*_{\text{ocm}}}{\partial \rho_c} \biggl|_{T,\kapeff^*,\Zeff^*}
\,\simeq\, \Zeff^* \quad \hbox{at}\quad \rho_c=\rho_c^*
\label{eq:check}
\ee
To compute the lhs of (\ref{eq:exe}), we may come back to the jellium model
which gives $\beta P^{\text{original}} \simeq \Zeff \rho_c$. 
In this expression, the effective charge may depend on the density,
but for salt-free cases, this dependence is at most logarithmic 
for $\rho_c\to 0$  \cite{Jellium}
and provides only a subdominant term to the compressibility, so that
\be
\frac{\partial \beta P^{\text{original}}}{\partial \rho_c} \biggl|_T \,\simeq \, \Zeff.
\ee
Evaluating this expression at $\rho_c=\rho_c^*$ where $\Zeff=\Zeff^*$, 
we recover Eq. (\ref{eq:check}). This not only illustrates the
identity (\ref{eq:exe}) but also the consistency of  the underlying  DLVO
potential.

%%%%%%%%%%%%%%%%%%%
\subsection{Situation with added salt}
When the suspension is dialyzed against a salt reservoir,
most of the technical analysis carried out earlier is still valid.
We consider a similar auxiliary system as in section \ref{ssec:back},
with effective screening length such that 
$\kapeff^{*2} > 4 \pi \lambda_B \Zeff^* \rho_c $ due to the screening
by salt ions \cite{rquejel}. The effective charge and
screening lengths are again chosen to coincide with those of a colloidal
system at a particular density $\rho_c^*$, but are otherwise
density independent. Equation (\ref{eq:exe}) still holds while 
$P_{\text{ocm}}^*$ is given by (\ref{eq:ocmvirial2}).
Neglecting again the integral on the rhs of (\ref{eq:ocmvirial2}), 
and inserting the resulting $P_{\text{ocm}}^*$ in (\ref{eq:exe}),
we obtain:
\be
\frac{\partial P^{\text{original}}}{\partial \rho_c} \biggl|_T  \, = \,
{\cal A}\, \,\frac{4 \pi \lambda_B \rho_c \Zeff^2}{\kapeff^{2}},
\label{eq:exebis}
\ee
where we have replaced $\Zeff^*$ by $\Zeff$ and $\kapeff^*$ by $\kapeff$
{\em after} computing 
the rhs of (\ref{eq:exe}). Here, the prefactor $\cal A$ reads
\be
{\cal A} \,=\,1+\frac{(\kapeff a)^2}{3(1+\kapeff a)^2}.
\ee

Is relation (\ref{eq:exebis}) compatible with $P_{\text{ocm}} \ll P = P^{\text{original}}$ ?
Neglecting $P_{\text{ocm}}$ (together with $\rho_c kT$) in (\ref{eq:P}), we have 
$P \simeq P_{\text{micro}}$ which in the jellium model is given by $\kapeff^2/(4\pi \lambda_B)$. With the help of \cite{rquejel}, we arrive at
\be
\frac{\partial P_{\text{micro}}}{\partial \rho_c} \biggl|_T  \, = \,
\frac{4 \pi \lambda_B \rho_c \Zeff^2}{\kapeff^{2}}.
\label{eq:pmicsalt}
\ee
Equations (\ref{eq:exebis}) and (\ref{eq:pmicsalt}) give the same result
provided ${\cal A}$ is close to unity, which means $\kapeff a < 1$.
We conclude here that omitting the colloidal contribution to the pressure,
$P_{\text{ocm}}$, is inconsistent when $\kapeff a > 1$.
It turns out however that ${\cal A}$ increases very mildly with $\kapeff a$
(e.g it is close to 1.2 for $\kapeff a=4$). A more precise discussion 
would require to consider the full rhs in (\ref{eq:ocmvirial2}), which is beyond the
scope of this paper. Finally, we note that in the salt-free case 
where $\kapeff a =3 \eta_c \Zeff \lambda_B/A$ with $\eta_c=4 \pi \rho_c a^3/3$ 
the colloidal volume fraction and $\Zeff \lambda_B/a$ on the order of 
10 for highly charged colloids, we have $\kapeff a <3$ and therefore
${\cal A}$ close to 1
even for packing fractions as high as 10\%. 

% We also note that one always has $\kapeff a > 3 \eta_c \Zeff \lambda_B/a$,
%with $\eta_c$ the colloid volume fraction. Given that $\Zeff \lambda_B/a$
%is typically of order 10 for highly charged colloids, neglecting 
%$P_{\text{ocm}}$ in the equation of state (\ref{eq:P}) requires roughly $\eta_c <0.05$.

%%%%%%%%%%%%%%%%%%%
\subsection{Summary}
We have seen that for a salt-free colloidal suspension, 
the colloidal contribution $P_{\text{ocm}}$
to the equation of state [as written in Eq. (\ref{eq:P})] 
is a negligible quantity.
This feature may easily be overlooked in a naive implementation
of the One Component Model, where only $P_{\text{ocm}}^*$,
the {\em bulk}
contribution to $P_{\text{ocm}}$, is computed.
The fact that $P_{\text{ocm}}^*$ is of the same order of magnitude
as the total pressure $P$ of the suspension, is not compatible
with the requirement $P_{\text{ocm}} \ll P_{\text{micro}} \simeq P$,
that has emerged as a central constraint in our analysis.
We have shown that no matter how large the system is,
surface effects that require the resolution of Poisson's
equation in the vicinity of a confining wall, contribute a large
amount to $P_{\text{ocm}}$. To zeroth approximation, these surface terms
cancel the bulk value $P_{\text{ocm}}^*$, so that 
one finally recovers $P_{\text{ocm}} \ll P$.

%%%%%%%%%%%%%%%%%%%%%%%%%%%%%%%%%%%%%%%%%%%%%
\section*{Acknowledgments}
It is a pleasure to thank Y. Levin 
for fruitful discussions.  R.C.P thanks PROMEP-Mexico and CONACyT
(grant 46373/A-1) for financial support.  J.D. acknowledges the
Marie-Curie fellowship MEIF-CT-2003-501789.  
This work has been supported in part by the NSF PFC-sponsored Center                          
for Theoretical Biological Physics (Grants No. PHY-0216576 and                                
PHY-0225630).  E.T. acknowledges the French ANR for an ACI.

%%%%%%%%%%%%%%%%%%%%%%%%%%%%%%%%%%%%%%%%%%%%


\begin{thebibliography}{10}
\bibitem{rque4} 
To be specific, the cell or jellium model pressure $P_{\text{micro}}$
gives an excellent approximation to the full primitive model pressure
$P$ as e.g. computed in \cite{Linse}. For a coupling $z^3
Z_c \lambda_B^2 / a^2 >1$, non mean-field effects become
important \cite{Netz,Levin}.
\bibitem{rquejel}
Within the jellium model in presence of an electrolyte,
one has $\kappa^4 = \kappa_{\text{res}}^4 + (4 \pi \lambda_B \rho_c \Zeff)^2$
where $\kappa_{\text{res}}$ is the inverse Debye length in the reservoir
against which the system is dialyzed. 
The precise form of the above relation
is however not required for the present discussion.
\end{thebibliography}
\end{document}